\documentclass[9pt,twocolumn,twoside]{pnas-new}

\templatetype{pnasresearcharticle} 

\title{History of art paintings through the lens of entropy and complexity}

\author[a]{Higor~Y.~D.~Sigaki}
\author[b,c,d,1]{Matja{\v z} Perc}
\author[a,1]{Haroldo~V.~Ribeiro}

\affil[a]{Departamento de F\'isica, Universidade Estadual de Maring\'a, Maring\'a, PR 87020-900, Brazil}
\affil[b]{Faculty of Natural Sciences and Mathematics, University of Maribor, Koro{\v s}ka cesta 160, SI-2000 Maribor, Slovenia}
\affil[c]{School of Electronic and Information Engineering, Beihang University, Beijing 100191, China}
\affil[d]{Complexity Science Hub, Josefst{\"a}dterstra{\ss}e 39, A-1080 Vienna, Austria}

\leadauthor{Sigaki}

\significancestatement{The critical inquiry of paintings is essentially comparative. This limits the number of artworks that can be investigated by an art expert in reasonable time. The recent availability of large digitized art collections enables a shift in the scale of such analysis through the use of computational methods. Our research shows that simple physics-inspired metrics that are estimated from local spatial ordering patterns in paintings encode crucial information about the artwork. We present numerical scales that map well to canonical concepts in art history and reveal a historical and measurable evolutionary trend in visual arts. They also allow us to distinguish different artistic styles and artworks based on the degree of local order in the paintings.}

\authorcontributions{HYDS, MP and HVR designed research, performed research, analyzed data, and wrote the paper.}
\authordeclaration{The authors declare that they have no conflict of interest.}
\correspondingauthor{\textsuperscript{1}To whom correspondence should be addressed. E-mail: hvr@dfi.uem.br, matjaz.perc@uni-mb.si.\\
This article contains supporting information online at \url{www.pnas.org/lookup/suppl/doi:10.1073/pnas.1800083115/-/DCSupplemental}}


\begin{abstract}
Art is the ultimate expression of human creativity that is deeply influenced by the philosophy and culture of the corresponding historical epoch. The quantitative analysis of art is therefore essential for better understanding human cultural evolution. Here we present a large-scale quantitative analysis of almost 140 thousand paintings, spanning nearly a millennium of art history. Based on the local spatial patterns in the images of these paintings, we estimate the permutation entropy and the statistical complexity of each painting. These measures map the degree of visual order of artworks into a scale of order-disorder and simplicity-complexity that locally reflects qualitative categories proposed by art historians. The dynamical behavior of these measures reveals a clear temporal evolution of art, marked by transitions that agree with the main historical periods of art. Our research shows that different artistic styles have a distinct average degree of entropy and complexity, thus allowing a hierarchical organization and clustering of styles according to these metrics. We have further verified that the identified groups correspond well with the textual content used to qualitatively describe the styles, and that the employed complexity-entropy measures can be used for an effective classification of artworks.
\end{abstract}

\doi{\url{www.pnas.org/cgi/doi/10.1073/pnas.1800083115}}

\begin{document}

\verticaladjustment{-2pt}

\maketitle
\thispagestyle{firststyle}
\ifthenelse{\boolean{shortarticle}}{\ifthenelse{\boolean{singlecolumn}}{\abscontentformatted}{\abscontent}}{}

\dropcap{P}hysics-inspired approaches have been successfully applied to a wide range of disciplines, including economic and social systems~\cite{mantegna1999introduction, wang2016statistical, perc2017statistical}. Such studies usually share the goal of finding fundamental principles and universalities that govern the dynamics of these systems~\cite{stanley1971phase}. The impact and popularity of this research has been growing steadily in recent years, in large part due to the unprecedented amount of digital information that is available about the most diverse subjects at an impressive degree of detail. This digital data deluge enables researchers to bring quantitative methods to the study of human culture~\cite{michel2011quantitative,dodds2015human,schich2014network}, mobility~\cite{gonzalez2008understanding,deville2016scaling} and communication~\cite{onnela2007structure,jiang2013calling,saramaki2014persistence}, as well as literature~\cite{hughes2012quantitative}, science production and peer review~\cite{kuhn2014inheritance,perc2013self,sinatra2015century,sinatra2016quantifying,balietti2016peer} at a scale that would have been unimaginable even a decade ago. A large-scale quantitative characterization of visual arts would be among such unimaginable research goals, not only because of data shortage, but also because the study of art is often considered to be intrinsically qualitative. Quantitative approaches aimed at the characterization of visual arts can contribute to a better understanding of human cultural evolution, as well as to more practical matters, such as to image characterization and classification.

While the scale of some current studies has changed dramatically, the use of quantitative techniques to the study of art has some precedent. Efforts can be traced back to the 1933 book ``Aesthetic Measure'' by the American mathematician Birkhoff~\cite{birkhoff1933aesthetic}, where a quantitative aesthetic measure is defined as the ratio between order (number of regularities found in an image) and complexity (number of elements in an image). However, the application of such quantitative techniques to the characterization of artworks is much more recent. Among the seminal works, we have the article by Taylor \textit{et al.}~\cite{taylor1999fractal}, where Pollock's paintings are characterized by an increasing fractal dimension over the course of his artistic career. This research article can be considered a landmark for the quantitative study of visual arts, inspiring many further applications of fractal analysis and related methods to determine the authenticity of paintings~\cite{jones2006fractal,taylor2006fractal,taylor2007authenticating,jones2009drip,de2016order}, to study the evolution of specific artists~\cite{boon2011artistic,alvarez2016fractal}, the statistical properties of particular paintings~\cite{pedram2008mona} and artists~\cite{taylor2004pollock,hughes2010quantification,shamir2012computer}, to study art movements~\cite{elsa2017topological}, and many other visual expressions~\cite{castrejon2003nasca,koch20101,montagner2016statistics}. The most recent advances of this emerging and rapidly growing field of research are comprehensively documented in several conference proceedings and special issues of scientific journals~\cite{Stork2008image,Stork2010image,Stork2011image,Stork2011image}, where contributions have been focusing also on artwork restoration tools, authentication problems, and stylometry assessment procedures.

To date, relatively few research efforts have been dedicated to study paintings from a large-scale art historical perceptive. In 2014, Kim \textit{et al.}~\cite{kim2014large} analyzed 29,000 images, finding that the color-usage distribution is remarkably different among historical periods of western paintings, and moreover, that the roughness exponent associated with the gray-scale representation of these paintings displays an increasing trend over the years. In a more recent work, Lee~\textit{et al.}~\cite{lee2017understanding} have analyzed almost 180 thousand paintings, focusing on the evolution of the color contrast. Among other findings, they have observed a sudden increase in the diversity of color contrast after 1850, and showed also that the same quantity can be used to capture information about artistic styles. Notably, there is also innovative research done by Manovich~\cite{ushizima2012cultural,manovich2015data,yazdani2017quantifying} concerning the analysis of large-scale datasets of paintings and other visual art expressions by means of the estimation of their average brightness and saturation.

However, except for the introduction of the roughness exponent, preceding research along similar lines has been predominantly focused on the evolution of color profiles, while the spatial patterns associated with the pixels in visual arts remain poorly understood. Here we present a large-scale investigation of local order patterns over almost 140 thousand visual artwork images that span several hundreds of years of art history. By calculating two complexity measures associated with the local order of the pixel arrangement in these artworks, we observe a clear and robust temporal evolution. This evolution is characterized with transitions that agree with different art periods. Moreover, the observed evolution shows that these periods are marked by distinct degrees of order and regularity in the pixel arrangements of the corresponding artworks. We further show that these complexity measures partially encode fundamental concepts of art history that are frequently used by experts for a qualitative description of artworks. In particular, the employed complexity measures distinguish different artistic styles according to their average order in the pixel arrangements, they enable a hierarchical organization of styles, and are also capable of automatically classifying artworks into artistic styles with significant accuracy.

\subsection*{Results}

Our results are based on a dataset comprising 137,364 visual artwork images (mainly paintings), obtained from the online visual arts encyclopedia WikiArt.org. This webpage is among the most significant freely available sources for visual arts. It contains artworks from over two thousand different artists, covering more than a hundred styles, and spanning a period of the order of a millennium. Each one of these image files has been converted into a matrix representation whose dimensions correspond to the image width and height, and whose elements are the average values of the shades of red, green and blue of the pixels in the RGB color space. For further details we refer to the Methods Section.

From this matrix representation of the artwork images, we calculate two complexity measures: the normalized permutation entropy $H$~\cite{bandt2002permutation} and the statistical complexity $C$~\cite{lopez1995statistical}. As described in Methods, both measures are evaluated from the ordinal probability distribution $P$, which quantifies the occurrence of the ordinal patterns among the image pixels at a local scale. Here we have estimated this distribution by considering sliding partitions of size $d_x=2$ by $d_y=2$ pixels (the embedding dimensions), leading to $(d_x d_y)!=24$ possible ordinal patterns. The value of $H$ quantifies the degree of disorder in the pixels arrangement of an image: values close to one indicate that pixels appear at random, while values close to zero indicate that pixels appear almost always in the same order. More regular images (such as those produced by the Minimalism) are expected to have small entropy values, while images exhibiting less regularity (such as Pollock's drip paintings) are characterized by large values of entropy. The statistical complexity $C$, in turn, measures the ``structural'' complexity present in an image~\cite{ribeiro2012complexity,zunino2016discriminating}: it is zero for both extremes of order and disorder in the pixels arrangement, and it is positive when an image presents more complex spatial patterns. The joint use of the values of $C$ versus $H$ as a discriminating tool give rise to a ``complexity-entropy plane''~\cite{lopez1995statistical,rosso2007distinguishing}, which is a technique that has proven to be useful in several applications. Herein the complexity-entropy plane is our approach of choice for quantifying the characteristics of different visual artworks. 

\subsection*{Evolution of Art}
A careful comparison of different artworks is one of the main methods used by art historians to understand whether and how art has evolved over the years. Works by Heinrich W\"olfflin~\cite{wolfflin1950principles} and Alois Riegl~\cite{riegl2004principles}, for example, can be considered fundamental in this regard. They have proposed to distinguish artworks from different periods through a few visual categories and qualitative descriptors. Visual comparison is undoubtedly a useful tool for evaluating artistic style. However, it is impractical to apply at scale. This is when computational methods show their greatest advantage. Nevertheless, in order to be useful it is important that derived metrics are still easily interpreted in terms of familiar and disciplinary relevant categories.

We note that the complexity-entropy plane partially (and locally) reflects W\"olfflin's dual concepts of linear vs. painterly and Riegl's dichotomy of haptic vs. optic artworks. According to W\"olfflin, ``linear artworks'' are composed of clear and outlined shapes, while in ``painterly artworks'' the contours are subtle and smudged for merging image parts and passing the idea of fluidity. Similarly, Riegl considers that ``haptic artworks'' depict objects as tangible discrete entities, isolated, and circumscribed, whereas ``optic artworks'' represent objects as interrelated in deep space by exploiting light, color, and shadow effects to create the idea of an open spatial continuum. The notions of order/simplicity vs. disorder/complexity in the pixel arrangements of images captured by the complexity-entropy plane partially encode these concepts. Images formed by distinct and outlined parts yield many repetitions of a few ordinal patterns, and consequently linear/haptic artworks are described by small values of $H$ and large values of $C$. On the other hand, images composed of interrelated parts delimited by smudged edges produce more random patterns, and accordingly, painterly/optic artworks are expected to yield larger values of $H$ and smaller values of $C$. It is also worth mentioning that W\"olfflin's and Riegl's dual concepts are limiting forms of representation that demarcate the scale of all possibilities~\cite{gaiger2002analysis}. In this regard, the continuum of $H$ and $C$ values may help art historians to grade this scale. 

\begin{figure}[!ht]
\begin{center}
\includegraphics[width=0.48\textwidth,keepaspectratio]{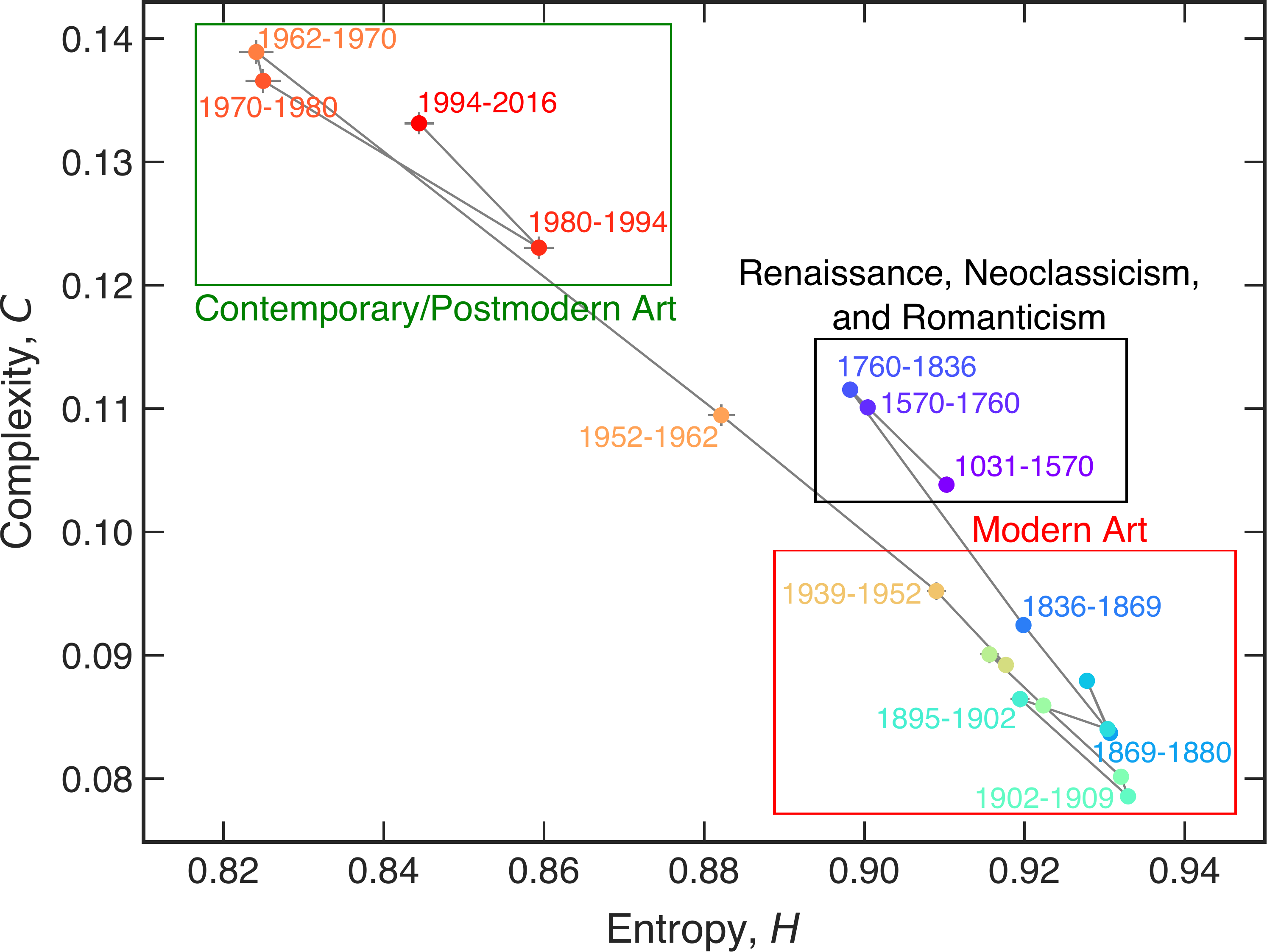}
\end{center}
\caption{\textbf{Quantifying the evolution of artworks through the history of art.} Temporal evolution of the average values of permutation entropy $H$ and statistical complexity $C$ (complexity-entropy plane). Each dot corresponds to the average values of $H$ and $C$ for a given time interval (shown in the plot). Error bars represent the standard error of the mean. The highlighted regions show different art periods (black: Renaissance, Neoclassicism, and Romanticism; red: Modern Art; green: Contemporary/Postmodern Art). We observe that the complexity-entropy plane correctly identifies different art periods and the transitions among them.}
\label{figure1}
\end{figure}

In this context, we ask whether the scale defined by $H$ and $C$ values is capable of unveiling any dynamical properties of art. To answer this question, we estimate the average values of $H$ and $C$ after grouping the images by date. Because the artworks are not uniformly distributed over time (see Methods), we have chosen time intervals containing nearly the same number of images in each time window. \autoref{figure1} shows the joint evolution of the average values of $C$ versus $H$ over the years (that is, the changes in the complexity-entropy plane), where a clear and robust (SI Appendix, Fig.~S1) trend is observed. This trajectory of $H$ and $C$ values shows that the artworks produced between the 9th and the 17th century are on average more regular/ordered than those created between the 19th and the mid 20th century. Also, the artworks produced after 1950 are even more regular/ordered than those from the two earlier periods. We observe further that the pace of changes in the complexity-entropy plane intensifies after the 19th century, a period that coincides with the emergence of several artistic styles (such as Neoclassicism and Impressionism), and also with the increase in the diversity of color contrast observed by Lee \textit{et al.}~\cite{lee2017understanding}. 

The three regions defined by the values of $H$ and $C$ correspond well with the main divisions of art history. The first period (black rectangle) corresponds to the Medieval Art, Renaissance, Neoclassicism, and Romanticism, which developed until the 1850s~\cite{danto1997after}. The second period (red rectangle) corresponds to the Modern Art, marked by the birth of Impressionism in the 1870s, and by the development of several avant-garde artistic styles (such as Cubism, Expressionism, and Surrealism) during the first decades of the 20th century. Finally, the latest period corresponds to the transition between Modern art and Contemporary/Postmodern Art. The specific date marking the beginning of the Postmodern period is still object of fierce debate among art experts~\cite{danto1997after}. Nevertheless, there is some consensus in that the Postmodern Art begins with the development of Pop Art in the 1960s~\cite{danto1997after}. 

By carrying the analogy between the complexity-entropy plane and the concepts of W\"olfflin and Riegl forward, the transition between the art produced before the modernism and Modern Art represents a change from linear/haptic to painterly/optic in the representation modes. This thus agrees with the idea that artworks from the Renaissance, Neoclassicism, and Romanticism usually represent objects rigidly distinguished from each other and separated by flat surfaces~\cite{wolfflin1950principles,kleiner_gardners_2013,hodge_history_2013}, while modern styles such as Impressionism, Fauvism, Pointillism, and Expressionism are marked by the use of looser and smudged brushstrokes in order to avoid the creation of pronounced edges~\cite{wolfflin1950principles,kleiner_gardners_2013,hodge_history_2013}. Intriguingly, the transition between Modern and Postmodern Art is marked by an even more intense and rapid change from painterly/optic to linear/haptic representation modes. This fact appears to agree with the postmodern idea of art as being instantly recognizable, made of ordinary objects, and marked by the use of large and well-defined edges (such as in Hard Edge Painting and Op Art artworks~\cite{kleiner_gardners_2013,hodge_history_2013}).

The conceptions of art history proposed by Wölfflin and Riegl consider that art develops through a change from the linear/haptic to the painterly/optic mode of representation, which agrees with the first transition observed in~\autoref{figure1}. However, for Riegl~\cite{blatt2014continuity}, this development occurs through a single and continuous process, while W\"olfflin has a cyclical conception of this transition that seems more consistent with the overall dynamical behavior of $H$ and $C$. On the other hand, this cyclical conception is not compatible with the local persistent behavior of the changes in the complexity-entropy plane. Indeed, recent studies of art historians, such as the work of Gaiger~\cite{gaiger2002analysis}, argue that neither of these conceptions hold when analyzing the entire development of art history. For Gaiger, the dual categories of W\"olfflin and Riegl should be treated as purely descriptive concepts and not linked to a particular change over time.

Another possibility of understanding the underlying mechanisms of the dynamical behavior unveiled by the complexity-entropy plane are evolutionary theories of art~\cite{boyd2005evolutionary,nadal2014evolutionary}. These recently proposed theories consider art from different perspectives, such as adaptation, a by-product of brain's complexity, sexual and natural selection aimed at sharing attention, and whose evolutionary contribution was to foster social cohesion and creativity. According to these theories, art history is driven by the interplay between audience preference and the artist desire to engage attention and expand these preferences. This feedback mechanism among artists and public would be responsible for propelling art towards its unprecedented degree of specialization, innovation, diversity, and could also explain what has driven artists and artistic movements to follow the historical path depicted in~\autoref{figure1}.

\subsection*{Distinguishing among artistic styles}

\begin{figure*}[!ht]
\begin{center}
\includegraphics[width=13.5cm,keepaspectratio]{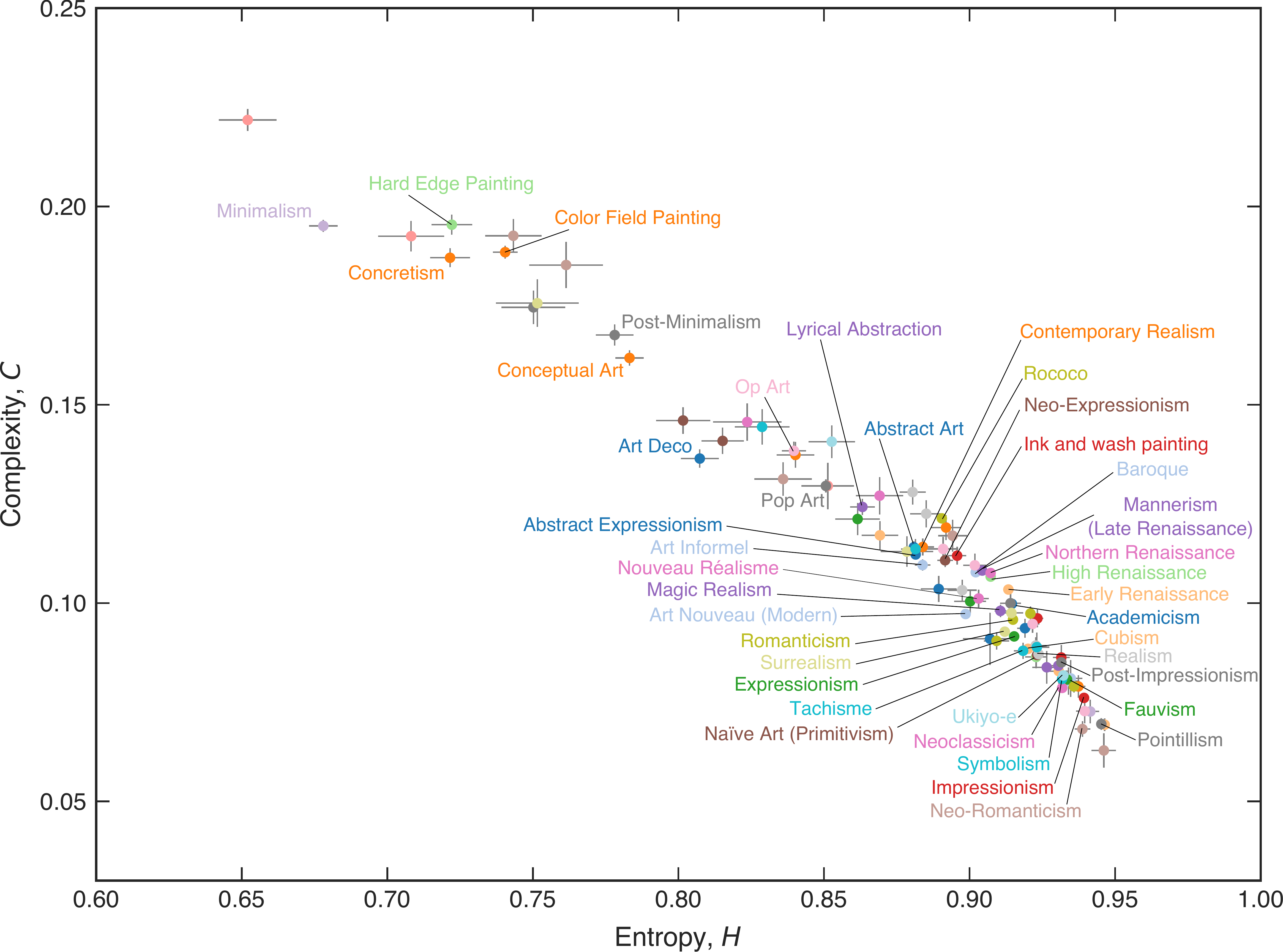}
\end{center}
\caption{\textbf{Distinguishing among different artistic styles with the complexity-entropy plane.} The colored dots represent the average values of $H$ and $C$ for each one of the 92 styles with more than 100 images in our dataset. Error bars represent the standard error of the mean. For better visualization, only 41 artistic styles with more than 500 images each are labeled (see SI Appendix, Fig.~S6 for all styles).}
\label{figure2}
\end{figure*}

We now ask whether the complexity-entropy plane is capable of discriminating among different artistic styles in our dataset. To do so, we calculate the average values of $H$ and $C$ after grouping the images by style. We also limit this analysis to the 92 styles having more than 100 images each (corresponding to $\approx$90\% of data -- see SI Appendix, Fig.~S5 for name and number of images of each one) to obtain reliable values for the averages. \autoref{figure2} shows that the artistic styles are spread over the complexity-entropy plane, and the average values of $H$ and $C$ are significantly different for the majority of the pairwise comparisons ($\approx$92\%, see SI Appendix, Fig.~S7). However, we also observe styles with statistically indistinguishable average values. 

We note further that the arrangement of styles is in agreement with the general trend in the average values of $H$ and $C$ over time in which most postmodern styles are localized in a region of smaller entropy and larger complexity values than modern styles (such as Expressionism, Impressionism, and Fauvism). This arrangement maps the different styles into a continuum scale whose extreme values partially reflect the dichotomy of linear/haptic vs. painterly/optic modes of representation. Among the styles displaying the highest values of $C$ and the smallest values of $H$, we find Minimalism, Hard Edge Painting, and Color Field Painting, which are all marked by the use of simple design elements that are well-delimited by abrupt transitions of colors~\cite{kleiner_gardners_2013,hodge_history_2013}. While styles displaying the smallest values of $C$ and the highest values of $H$ (such as Impressionism, Pointillism, and Fauvism) are characterized by the use of smudged and diffuse brushstrokes, and also by blending colors in order to avoid the creation of sharp edges~\cite{kleiner_gardners_2013,hodge_history_2013}.

\subsection*{Hierarchical structure of artistic styles}
\begin{figure*}[!ht]
\begin{center}
\includegraphics[width=1.\textwidth,keepaspectratio]{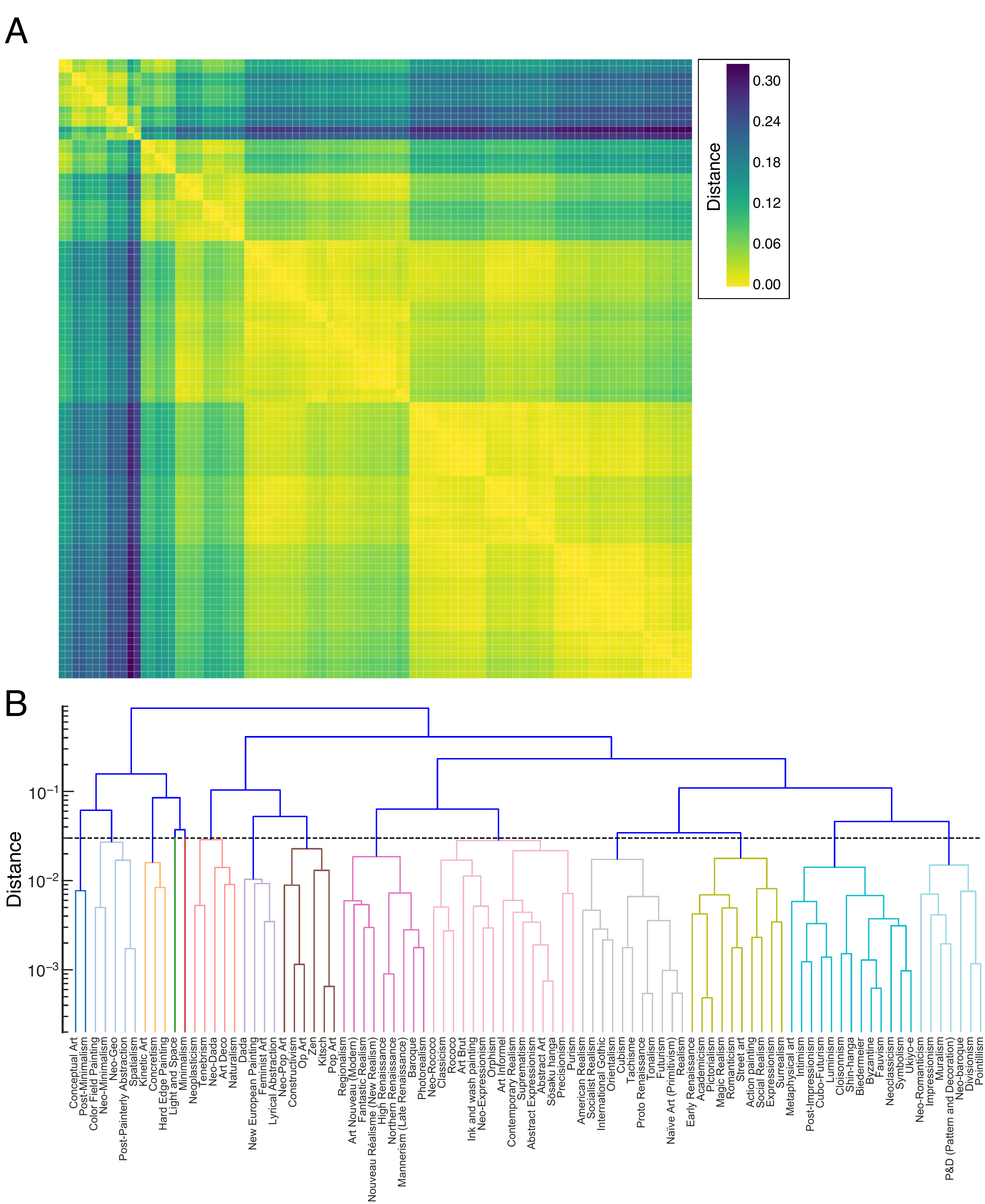}
\end{center}
\caption{\textbf{Hierarchical organization of artistic styles.} (A) Matrix plot of the Euclidean distance in the complexity-entropy plane between every pair of styles. (B) Dendrogram representation of the distance matrix obtained by applying the minimum variance method proposed by Ward~\cite{ward1963hierarchical}. The 14 groups of styles indicated by the colored branches are obtained by cutting the dendrogram at the threshold distance 0.03. This value maximizes the silhouette coefficient~\cite{rousseeuw1987silhouettes} (see Methods Section~\ref{meth_silhouette} and~SI Appendix, Fig.~S8) and it is thus a ``natural'' number for defining the number of clusters in our dataset. The order of rows and columns in the matrix plot is the same as used in the dendrogram.}
\label{figure3}
\end{figure*}

The values of $H$ and $C$ capture the degree of similarity among artistic styles regarding the local ordering of image pixels. This fact enables us to test for a possible hierarchical organization of styles with respect to this local ordering. To do so, we have considered the Euclidean distance between a pair of styles in the complexity-entropy plane as a dissimilarity measure between them. Thus, the closer the distance between two artistic styles, the more significant is the similarity between them, whereas pairs of styles separated by large distances are considered more dissimilar from each other. \autoref{figure3}(A) shows the matrix plot of these distances, where we qualitatively observe the formation of style groups.

To investigate the clustering between artistic styles systematically, we employ the minimum variance method proposed by Ward~\cite{ward1963hierarchical} to construct a dendrogram representation of the distance matrix. This method is a hierarchical clustering procedure that uses the within-cluster variance as the criterion for merging pairs of clusters. \autoref{figure3}(B) depicts this dendrogram, unveiling an intricate relationship among the artistic styles in our dataset. By maximizing the silhouette coefficient~\cite{rousseeuw1987silhouettes} (as described in Methods and SI Appendix, Fig.~S8), we find that $0.03$ is the optimal threshold distance that maximizes the cohesion and separation among the clusters of styles. This threshold distance yields 14 groups of styles indicated by the different colors in~\autoref{figure3}(A)~and~(B). 

These groups partially reflect the temporal localization of different artistic styles and their evolution reported in~\autoref{figure1}. In particular, several styles that emerged together or close in time are similar regarding the local arrangement of pixels and thus belong to the same group. For instance, the first five groups of~\autoref{figure3}(B) contain mainly Postmodern styles. On the other hand, these groups and their hierarchical structure organize the styles regarding their mode of representation in the scale delimited by the dichotomy of linear/haptic vs. painterly/optic. This fact is more evident when examining groups in both extremes of order and regularity in the complexity-entropy plane. The right-most group of~\autoref{figure3}(B), for example, contains styles that employ relatively small brush strokes and avoid the creation of sharp edges. This fact is particularly evident in artworks of Impressionism, Pointillism, Divisionism, but it is also evident in Neo-Baroque and Neo-Romanticism, and in the works of muralists (such as David Siqueiros and Jos\'e Orozco), as well as in the abstract paintings of P\&D (Pattern and Decoration). While devoted to patterning paintings (such as printed fabrics), P\&D is considered a ``reaction'' to Minimalism and Conceptual Art (which are located in the other extreme of the complexity-entropy plane) that avoids restrained compositions by means of a subtle modulation of colors as in the works of Robert Zakanitch, who is considered one of the founders of P\&D~\cite{swartz2007pattern}. As we move to groups characterized by high complexity and low entropy, we observe the clustering of styles marked by the presence of sharped edges and very contrasting patterns, usually formed by distinct parts isolated or combined with unrelated materials. That is the case for the group containing Op Art, Pop Art, and Constructivism, but also for the group formed by Kinetic Art, Hard Edge Painting, and Concretism~\cite{kleiner_gardners_2013,hodge_history_2013}.

We can also verify the meaningfulness of these groups by comparing the clustering of~\autoref{figure3}(B) with an approach based on the similarities among the textual content of the Wikipedia pages of each artistic style. To do so, we have obtained the textual content of these webpages and extracted the top-100 keywords of each one by applying the term frequency-inverse document frequency (TF-IDF) approach~\cite{chowdhury2010introduction}. We consider the inverse of $1$ plus the number of shared keywords between two styles as a measure of similarity between them. Thus, styles having no common keywords are at the maximum ``distance'' of 1, while styles sharing several keywords are at a closer ``distance''. 

By employing a similar hierarchical clustering procedure as the one used in~\autoref{figure3}, we obtain 24 clusters of artistic styles from the Wikipedia text analysis (SI Appendix, Fig.~S9). This number of clusters is much larger than the 14 clusters obtained from the complexity-entropy plane. However, both clustering approaches share similarities, which can be quantified by employing the clustering evaluation metrics homogeneity \textit{h}, completeness \textit{c}, and $v$-measure~\cite{rosenberg2007v}. Perfect homogeneity ($h=1$) implies that all clusters obtained from the Wikipedia texts contain only styles belonging to the same clusters obtained from the complexity-entropy plane. On the other hand, perfect completeness ($c=1$) implies that all styles belonging to the same cluster obtained from the complexity-entropy plane are grouped in the same cluster obtained from the Wikipedia texts. The $v$-measure is the harmonic mean between $h$ and $c$, that is, $v=2 h c/(h+c)$. Our results yield $h=0.49$, $c=0.40$, and $v=0.44$, which are values significantly larger than those obtained from a null model where the number of shared keywords is randomly chosen from a uniform distribution between 0 and 100 ($h_{\text{rand}}=0.42\pm0.02$, $c_{\text{rand}}=0.35\pm0.01$, and $v_{\text{rand}}=0.38\pm0.01$ -- average values over 100 realizations). Therefore, the similarities between both clustering approaches cannot be explained by chance. This result indicates that in spite of the very local character of our complexity measures, the values of $H$ and $C$  reflect the meaning of some keywords used for describing artistic styles. 

\subsection*{Predicting artistic styles}\label{res:predicting}

\begin{figure}[ht]
\begin{center}
\includegraphics[width=0.5\textwidth]{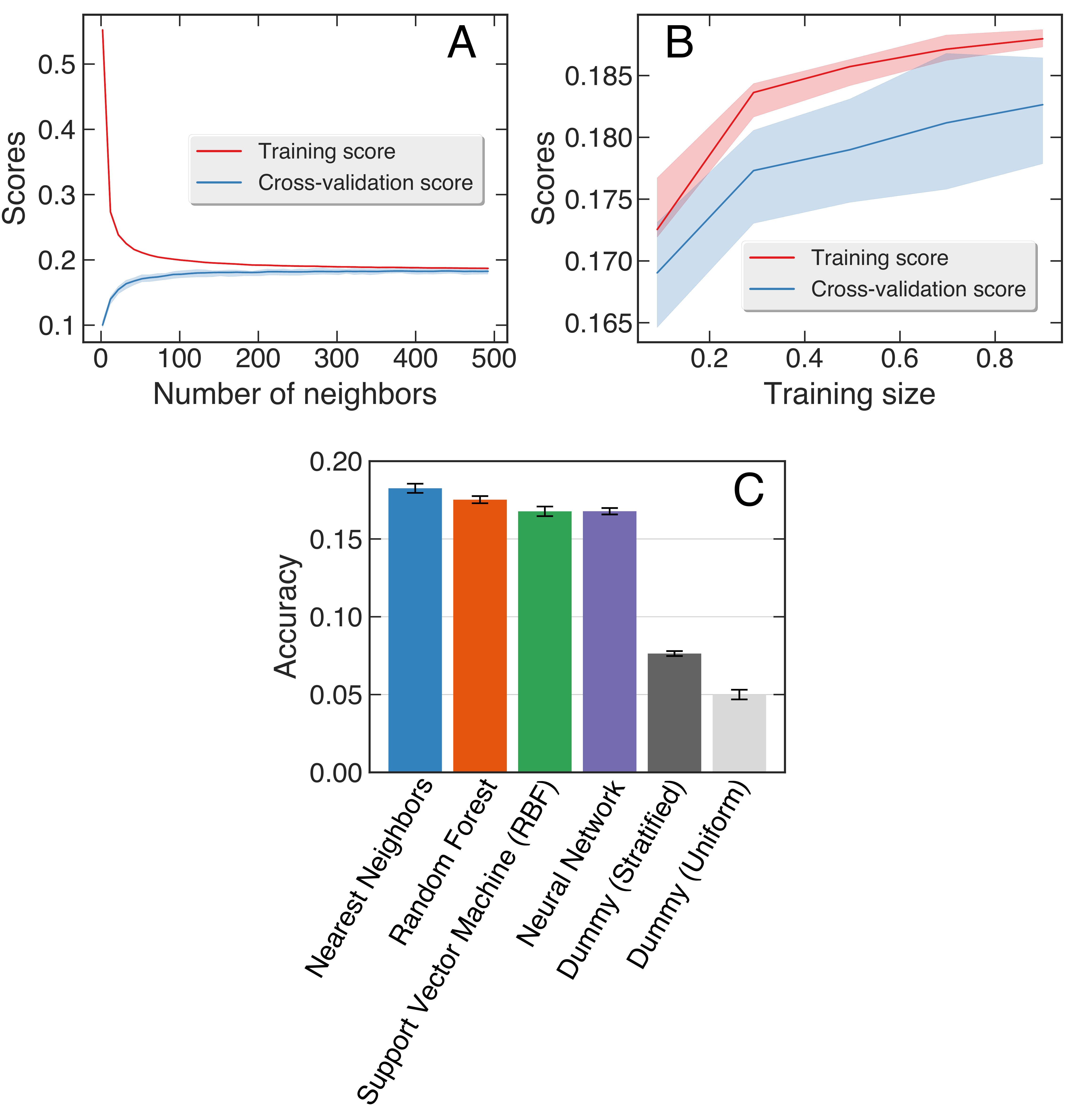}
\end{center}
\caption{\textbf{Predicting artistic styles with statistical learning algorithms.} (A) Training and cross-validation scores of the nearest neighbors algorithm as a function of the number of neighbors. We note that the algorithm underfits the data for values of the number of neighbors smaller than $\approx$250, but there is no significant accuracy improvement for larger values nor for overfitting up to 500 neighbors. (B) Learning curve, that is the training and cross-validation scores, as a function of the training size (fraction of the whole data) for the nearest neighbors algorithm with the number of neighbors equal to 400. We observe no significant improvement in the cross-validation scores when more than 50\% of the data is used to train the model. In both plots, the shaded regions are 95\% confidence intervals obtained with a 10-fold cross-validation splitting strategy. (C) Comparison between four different statistical learning algorithms (nearest neighbors, random forest, support vector machine, and neural network; see SI Appendix, Fig.~S10 for details of the parameters for each algorithm) as well as the null accuracy obtained from two ``dummy'' classifiers (stratified: generates random predictions respecting the style distributions; uniform: predictions are uniformly random). The four classifiers have similar accuracy ($\approx$18\%), and they all significantly outperform the ``dummy'' classifiers. These results are based on the 20 styles with more than 1500 images each, although similar results are obtained when including others styles as well (SI Appendix, Fig.~S11).}
\label{figure4}
\end{figure}

Another possibility of quantifying the information encoded by the values of $H$ and $C$ is trying to predict the style of an image based only on these two values. To do so, we have implemented four well-known machine learning algorithms~\cite{hastie2013elements,muller2016introduction} (nearest neighbors, random forest, support vector machine, and neural network -- see Methods for details) for the classification task of predicting the style of images for all 20 styles that contain more than 1500 artworks each. For each method, we estimate the validation curves for a range of values of the main parameters of the algorithms with a stratified $n$-fold cross-validation~\cite{hastie2013elements} strategy with $n=10$. \autoref{figure4}(A) shows the validation curves for the $k$-nearest neighbors as a function of the number of neighbors. We note that this method underfits the data if the number of neighbors is smaller than $\approx$250. Conversely, the cross-validation score saturates at $\approx$0.18 if the neighbors are 300 or more, and there is no overfitting up to 500 neighbors. Another relevant issue for statistical learning is related to the amount of data necessary to properly train the model. To investigate this, we again employ a stratified $n$-fold cross-validation strategy with $n=10$ for estimating the learning curves. \autoref{figure4}(B) shows the training and cross-validation scores for the $k$-nearest neighbors, where we observe that both scores increase with training size. However, this enhancement is very small when more than $\approx$50\% of the data is used for training the model. SI Appendix, Fig.~S10 shows results analogous to those presented in~\autoref{figure4} as obtained with the other three machine learning algorithms.

By combining the previous analysis with a grid-search algorithm, we determine the best combination of parameters enhancing the performance of each statistical learning method. \autoref{figure4}(C) shows that the four algorithms display similar performances, all exhibiting accuracies close to $18$\%. We have further compared these accuracies with those obtained from two ``dummy'' classifiers. In the stratified classifier, style predictions are generated by chance but respecting the distribution of styles, while predictions are drawn uniformly at random when using the uniform classifier. The results in \autoref{figure4}(C) show that all machine learning algorithms have a significantly larger accuracy than is obtained by chance. This result thus confirms that the values of $H$ and $C$ encode important information about the style of each artwork. Nevertheless, the achieved accuracy is quite modest for practical applications. Indeed, there are other approaches that are more accurate. For instance, Zujovic \textit{et al.}~\cite{zujovic2009classifying} achieved accuracies of $\approx$70\% in a classification task with 353 paintings from 5 styles, and Agarwal \textit{et al.}~\cite{agarwal2015genre} reported an accuracy of $\approx$60\% in a classification task with 3000 paintings from 10 styles. However, our results cannot be directly compared with those works since they employ a much smaller dataset with fewer styles and several image features, while our predictions are based only on two features. Our approach represents a severe dimensionality reduction since images with roughly one million pixels are represented by two numbers related to the local ordering of the image pixels. In this context, an accuracy of $18$\% in a classification with 20 styles and more than 100 thousand artworks is not negligible. Moreover, the local nature of $H$ and $C$ makes these complexity measures very fast, easy to parallelize, and scalable from the computational point of view. Thus, in addition to showing that the complexity-entropy plane encodes important information about the artistic styles, we believe that the values of $H$ and $C$, combined with other image features, are likely to provide better classification scores.

\subsection*{Discussion and Conclusions}
We have presented a large-scale characterization of a dataset composed of almost 140 thousand artwork images that span the latest millennium of art history. Our analysis is based on two relatively simple complexity measures (permutation entropy $H$ and statistical complexity $C$) that are directly related to the ordinal patterns in the pixels of these images. These measures map the local degree of order of these artworks into a scale of order-disorder and simplicity-complexity that locally reflects the qualitative description of artworks proposed by W\"olfflin and Riegl. The limits of this scale correspond to two extreme modes of representation proposed by these art historians, namely to the dichotomy between linear/haptic ($H\approx0$ and $C\approx0$) vs. painterly/optic ($H\approx1$ and $C\approx0$). 

By investigating the dynamical behavior of the average values of the employed complexity measures, we have found a clear and robust trajectory of art over the years in the complexity-entropy plane. This trajectory is characterized by transitions that agree with the main periods of art history. These transitions can be classified as linear/haptic to painterly/optic (before and after the Modern Art) and painterly/optic to linear/haptic (the transition between Modern and Postmodern Art), showing that each of these historical periods has a distinct degree of entropy and complexity. While W\"olfflin's conception of art history in terms of a cyclical transition between linear and painterly does not withstand the local time persistence in the values of $H$ and $C$ nor the critical scrutiny of Gaiger~\cite{gaiger2002analysis} and other contemporary art historians, it is quite consistent with the global evolution depicted in the complexity-entropy plane. For W\"olfflin, the transition from linear to painterly is governed by a ``natural law in the same way as physical growth'' and ``to determine this law would be a central problem, the central problem of history of art'' (page 17 of~\cite{wolfflin1950principles}). However, the return to the linear ``lies certainly in outward circumstances'' (page 233 of~\cite{wolfflin1950principles}) and, in the context of~\autoref{figure1}, it is not difficult to envisage that the transition from Modern to Postmodern was driven by the end of World War II, the event that usually marks the beginning of postmodernism in history books. 

In addition to unveiling this dynamical aspect of art, the values of $H$ and $C$ are capable of distinguishing between different artistic styles according to the average degree of entropy/complexity in the corresponding artworks. We emphasize that the location of each style in the complexity-entropy plane partially reflects the duality linear/haptic vs. painterly/optic, and thus can be considered as a ruler for quantifying the use of these opposing modes of representation. Also, the distances between pairs of styles in the complexity-entropy plane represent a similarity measure regarding these art history concepts. By using these distances, we find that different styles can be hierarchically organized and grouped according to their position on the plane. We have verified that these groups reflect well the textual content of Wikipedia pages used for describing each style, and they also reflect some similarities among them, in particular regarding the presence of soft/smudged/diffuse or well-defined/sharp/abrupt transitions. We have further quantified the amount of information encoded in these complexity measures by means of a classification task in which the style of an image is predicted based solely on the values of $H$ and $C$. The obtained success rate of approximately 18\% outperforms dummy classifiers, in turn showing that these two measures carry meaningful information about artwork style. 

Since our two complexity measures are based entirely on the local scale of an artwork, they of course cannot capture all the uniqueness and complexity of art. However, our results nevertheless demonstrate that simple physics-inspired metrics can be connected to concepts proposed by art historians, and more importantly, that these measures do carry relevant information about artworks, their style, and evolution. In the context of W\"olfflin's metaphor about the evolution of art: ``A closer inspection certainly soon shows that art even here did not return to the point at which it once stood, but that only a spiral movement would meet the facts.'' (page 234 of~\cite{wolfflin1950principles}), we may consider the complexity-entropy plane as one of the possible projections of W\"olfflin's spiral. 

\matmethods{

\subsection*{Data}\label{meth_data}

The digital images used in this study were obtained from the visual arts encyclopedia WikiArt~\cite{WikiArt}, which is one of the largest online and freely available datasets of visual artworks available to date. By crawling the web pages of WikiArt in August of 2016, we have downloaded 137,364 digitalized images and metadata related to each artwork, such as painter (there are 2391 different artists), date, and artistic style (\textit{e.g.}, Impressionism, Surrealism, and Baroque). The style labels provided by WikiArt.org are generated and collaboratively maintained by the users of that webpage. For the analysis of the temporal evolution, we have excluded all images whose composition dates were not specified (33,724 files). \autoref{figure5}(A) depicts the number of images per year in our dataset, where we observe that these artworks were created between the years 1031 and 2016. \autoref{figure5}(B) shows that the cumulative fraction of artworks in our dataset is well approximated by an exponential growth with the characteristic time equal to $\tau=111\pm1$ years. Consequently, the cumulative number of artworks is doubling every 77 years. Also, more than 50\% of these artworks were produced after the first decade of the 20th century -- a period that is marked by the development of a large variety of different art movements.

\begin{figure}
\begin{center}
\includegraphics[width=0.33\textwidth]{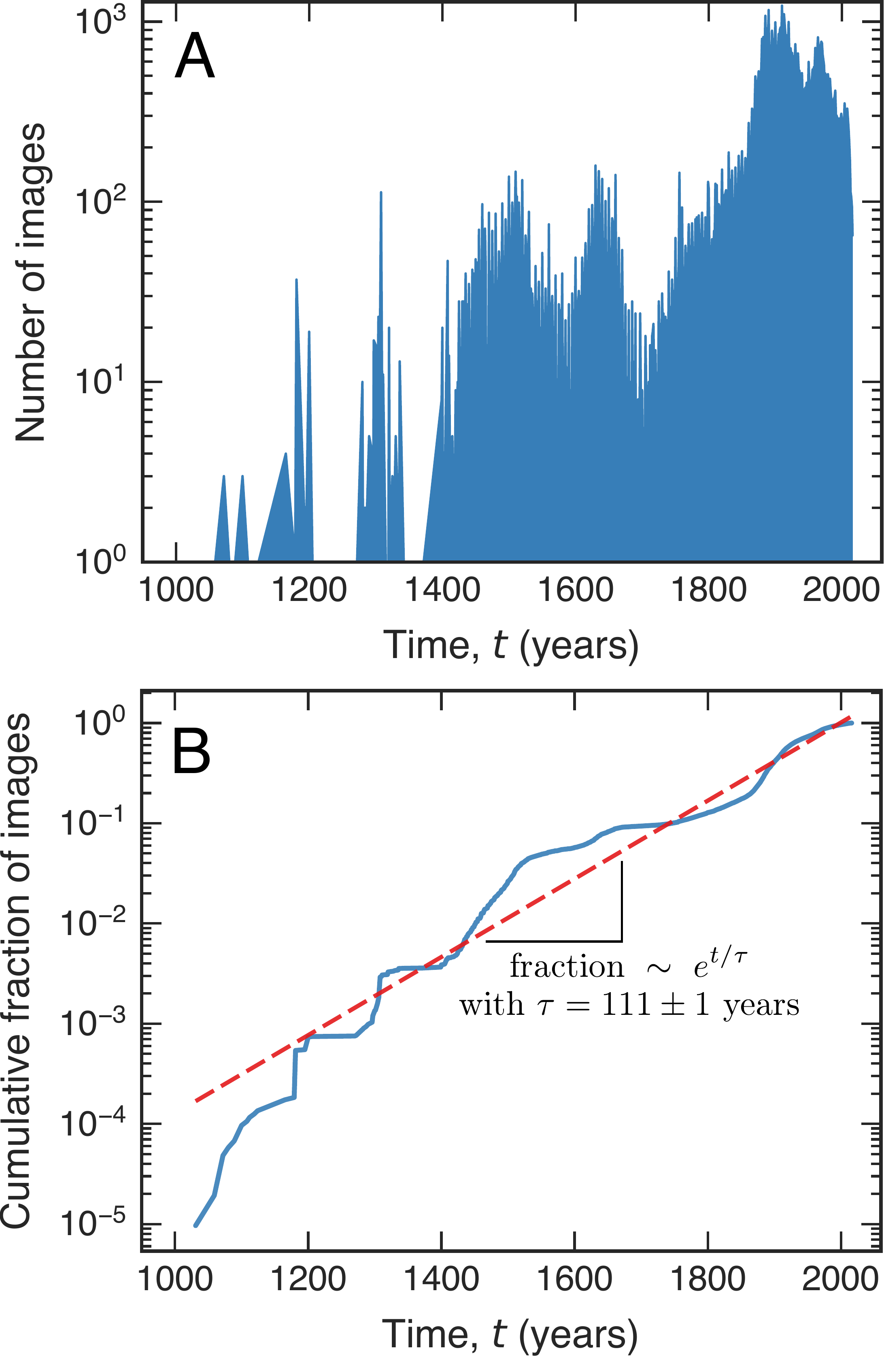}
\end{center}
\caption{\textbf{Distribution of artwork images over the years.} (A) The number of images per year in our dataset. (B) Cumulative fraction of artworks over the years (blue curve) on a log-linear scale. This fraction [$f(t)$] is approximated by an exponential growth [$f(t) \propto \exp(t/\tau)$] with characteristic time equal to $\tau=111\pm1$ years. We observe that most artworks were produced after the onset of the 20th century, in particular, more than 50\% were painted after the year 1912.}
\label{figure5}
\end{figure}

\subsection*{Matrix representation of image files}\label{meth_images}
All image files are in JPEG format with 24 bits per pixel (8 bits for red, green and blue colors in the RGB ``color space''), meaning that each pixel of the image is characterized by 256 shades of red, green, and blue, which in total allows $256^3=16,777,216$ color variations. For practical purposes, an image file can be thought of as a three-layer matrix of dimensions $n_x$ (the image width) per $n_y$ (the image height), where the layers correspond to the color channels and the elements (ranging from 0 to 255) represent the color intensity. For our analysis, we have calculated the average value of the three color shades of each pixel, yielding a simple matrix for each image file. This approach is similar to the usual gray-scale transformation of images, except that the average over the three color channels is usually weighted by different values. One of the most common weighting values defines the gray-scale reflectance or luminance~\cite{scikit-image} and corresponds to calculating $0.2125 R + 0.7154 G + 0.0721 B$, where $R$, $G$ and $B$ stand for the shade intensities of red, green, and blue, respectively. These weighting values are often chosen to mimic the color sensibility of the human eye, but our results are remarkably robust against different weighting choices. For example, the Pearson linear correlation between the values of $H$ calculated with the usual gray-scale transformation and the simple average is $0.989$ and $0.992$ for the values of $C$ -- SI Appendix, Fig.~S2. We have therefore resorted to using the simple average value.

\subsection*{The complexity-entropy plane}\label{meth_plane}
By using the matrix representation of all images, we calculate the normalized permutation entropy $H$ and statistical complexity $C$ for each one. This technique was originally proposed for characterizing time series~\cite{bandt2002permutation,rosso2007distinguishing}, and only recently it has been generalized to use with higher dimensional data such as images~\cite{ribeiro2012complexity,zunino2016discriminating}. Here we shall present this technique through a simple example (for a more formal description we refer to original articles). Let the matrix
\begin{equation*}
A =
\begin{bmatrix}
6 & 0 & 2 \\
4 & 5 & 2 \\
6 & 7 & 4
\end{bmatrix}
\end{equation*}
represent a hypothetical image of size $3\times3$. The first step is to define sliding sub-matrices of size $d_x$ per $d_y$, where these values are called embedding dimensions (the only parameters of the method). By choosing $d_x=d_y=2$, we have the following four partitions
\begin{equation*}
A_1 =
\begin{bmatrix}
6 & 0\\
4 & 5
\end{bmatrix},~~
A_2 =
\begin{bmatrix}
0 & 2\\
5 & 2
\end{bmatrix},~~
A_3 =
\begin{bmatrix}
4 & 5\\
6 & 7
\end{bmatrix},~\text{and}~~
A_4 =
\begin{bmatrix}
5 & 2\\
7 & 4
\end{bmatrix}.
\end{equation*}
The next step is to investigate the ordinal patterns of occurrence of the elements of these sub-matrices. By letting $a_0$, $a_1$, $a_2$, and $a_3$ represent the elements of these matrices line by line, we have that $A_1$ is described by the ordinal pattern $\Pi_1=(1,2,3,0)$ since $a_1<a_2<a_3<a_0$, where $\Pi_1$ represents the permutation that sorts the elements of $A_1$ in ascending order. Similarly, $A_3$ is described by $\Pi_3=(0,1,2,3)$ (since $a_0<a_1<a_2<a_3$) and $A_4$ by $\Pi_4=(1,3,0,2)$ (since $a_1<a_3<a_0<a_2$). In case of draws, we keep the occurrence order of the elements $a_0$, $a_1$, $a_2$, and $a_3$. Thus, $A_2$ is described by $\Pi_2=(0,1,3,2)$ since $a_0<a_1\leq a_3<a_2$ and because $a_1$ precedes $a_3$. From all ordinal patterns associated with $A$ for a given $d_x$ and $d_y$, we estimate the probability distribution of finding each permutation from its relative frequency of occurrence. In our example, from all 24 possible permutations [that is, $(d_x d_y)!$], four have appeared just once, and thus their probabilities are $1/4$, while all other permutations have zero probability. Therefore, the probability distribution of the ordinal patterns associated with $A$ is $P\!\!=\{1/4, 1/4, 1/4, 1/4,0,\dots,0\}$. We have omitted the specification of each permutation in $P$ because the order of its elements is irrelevant for the described procedure.

Having the probability distribution $P=\{p_i;~i=1,\dots,n\}$, we calculate the normalized Shannon entropy
\begin{equation}
H(P) = \frac{1}{\ln(n)} \sum_{i=0}^{n} p_i \ln (1/p_i)\,,
\end{equation}
where $n=(d_x d_y)!$ is the number of possible permutations and $\ln(n)$ corresponds to the maximum value of the Shannon entropy $S(P)=\sum_{i=0}^{n} p_i \ln (1/p_i)$, that is, when all permutations are equally likely to occur ($p_i=1/n$). The value of $H$ quantifies the degree of ``disorder'' in the occurrence of the pixels of an image represented by the matrix $A$. We have $H \approx 1$ if the pixels appear in random order, and $H \approx 0$ if they always appear in the same order.

In spite of the value of $H$ being a good measure of randomness, it cannot adequately capture the degree of structural complexity present in $A$~\cite{zunino2016discriminating}. Because of that, we further calculate the so-called statistical complexity~\cite{lopez1995statistical,lamberti2004intensive,martin2006generalized}
\begin{equation}
C(P)=\frac{D(P,U) H(P)}{D^*}\,,
\end{equation}
where $D(P,U)$ is a relative entropic measure (the Jensen-Shannon divergence) between\linebreak \mbox{$P=\{p_i;~i=1,\dots,n\}$} and the uniform distribution $U=\{u_i=1/n;~i=1,\dots,n\}$ defined as
\begin{equation}
D(P,U) = S\left(\frac{P+U}{2}\right) - \frac{S(P)}{2} - \frac{S(U)}{2}\,,
\end{equation}
where $\frac{P+U}{2} = \{\frac{p_i+1/n}{2},~i=1,\dots,n\}$ and
\begin{equation}
D^*\!\!=\!\max_{P}D(P,U)=-\frac{1}{2}\left[\frac{n+1}{n}\ln(n+1) + \ln(n) - 2 \ln(2n)\right]\!\!,
\end{equation}
is a normalization constant (obtained by calculating $D(P,U)$ when just one component of $P$ is equal to one and all others are zero). The quantity $D(P,U)$ is zero when all permutations are equally likely to happen, and it is larger than zero if there are privileged permutations. Thus, $C(P)$ is zero in both extremes of order (\mbox{$P=\{p_i=\delta_{1,i};~i=1,\dots,n\}$}) and disorder (\mbox{$P=\{p_i=1/n;~i=1,\dots,n\}$}). However, $C(P)$ is not a trivial function of $H(P)$. Namely, for a given $H(P)$ there exists a range of possible values for $C(P)$~\cite{martin2006generalized,rosso2007distinguishing,zunino2016discriminating}, from which $C(P)$ quantifies the existence of structural complexity and provides additional information that is not carried by the value of $H(P)$. Mainly because of this fact, Rosso \textit{et al.}~\cite{rosso2007distinguishing} proposed to use a diagram of $C(P)$ versus $H(P)$ -- a representation space that is called the complexity-entropy plane~\cite{lopez1995statistical}.

We estimate the values of $H(P)$ and $C(P)$ for all 137,364 images by considering the embedding dimensions $d_x=d_y=2$. As we previously mentioned, these values are the only ``tuning'' parameters of the permutation complexity-entropy plane. However, the choice for the embedding dimensions is not completely arbitrary as the condition $(d_x d_y)!\ll n_x n_y$ must hold in order to obtain a reliable estimate of the ordinal probability distribution $P$~\cite{ribeiro2012complexity}. In our dataset, the average values of image length $n_x$ and height $n_y$ are both close to 900 pixels (SI Appendix, Fig.~S3), thus practically limiting our choice to $d_x=d_y=2$.

\subsection*{Independence of $H$ and $C$ on image dimensions}\label{meth_dimensions}
The image files obtained from WikiArt do not have the same dimensions. SI Appendix, Fig.~S3 shows that image width and height have a similar distribution, with average values equal to $895$ pixels for width and 913 pixels for height. Also, 95\% of the images have width between 313 and 2491 pixels, and height between 323 and 2702 pixels. Because of these different dimensions, we have tested whether the values of $H(P)$ and $C(P)$ display any bias as a result. This is an important issue since we expect the values of $H(P)$ and $C(P)$ to reflect the characteristics of an image, not its dimensions. SI Appendix, Fig.~S4 shows scatter plots of the values of $H(P)$ and $C(P)$ versus the square root of image areas (that is, $\sqrt{n_x n_y}$) at several different scales, where any visual relationship is observed. The Pearson linear correlation is very low ($\approx$0.05) for both relationships. We also estimate the maximal information coefficient (MIC)~\cite{Reshef}, a non-parametric coefficient that measures the association between two variables, even if they are correlated in a nonlinear fashion. The MIC value is very small ($\approx$0.07) for both relationships. Thus, we conclude that the values of $H(P)$ and $C(P)$ are not affected or biased by image dimensions.

\subsection*{Finding the number of clusters with the silhouette coefficient}\label{meth_silhouette}
The hierarchical organization of artistic styles presented in~\autoref{figure3}B enables the determination of clusters of styles. To do so, we must choose a threshold distance for which styles belong to different clusters. The number of clusters naturally depends on this choice. A way of determining an optimal threshold distance is by calculating the silhouette coefficient~\cite{rousseeuw1987silhouettes}. This coefficient evaluates both the cohesion and the separation of data grouped into clusters. The silhouette coefficient is defined by the average value of
\begin{equation}
s_i = \frac{b_i-a_i}{\max(a_i,b_i)}\,,
\end{equation}
where $a_i$ is the average intra-cluster distance and $b_i$ is the average nearest-cluster distance for the $i$-th datum. By definition, the silhouette coefficient ranges from $-1$ to $1$, and the higher its value the better the clustering configuration. SI Appendix, Fig.~S8 shows the silhouette coefficient as a function of the threshold distance used for determining the clusters in \autoref{figure3}B. We observe that the coefficient displays a maximum (of $0.57$) when the threshold distance is 0.03. Thus, this threshold distance is the one that maximizes the cohesion and separation among the artistic styles. By using this value, we find 14 different groups of artistic styles shown in~\autoref{figure3}. Also, a similar approach yields the 24 different clusters that are associated with the similarities among the Wikipedia pages of the styles reported in SI Appendix, Fig.~S9.

\subsection*{Implementation of machine learning algorithms}\label{meth_machine_learning}
All machine learning algorithms employed for predicting the artistic styles from an image are implemented by using functions of the Python scikit-learn library~\cite{scikit-learn}. For instance, the function \texttt{sklearn.neighbors.KNeighborsClassifier} implements the $k$-nearest neighbors. In statistical learning, a classification task involves inferring the category of an object by using a set of explanatory variables or features associated with this object, and the knowledge of other observations (the training set) in which the categories of the objects are known. For the results presented in the main text, we have included 20 different styles that each have more than 1500 images in our analysis (see SI Appendix, Fig.~S5 for names). However, similar results are obtained by considering a larger number of styles. For example, the overall accuracy of different learning algorithms is approximately 13\% if we consider all the styles with more than 100 images each (SI Appendix, Fig.~S11) (compared to approximately 18\% if only 20 styles are considered).

Thus, our classification task involves identifying the artistic style (the categories) of an image from its entropy $H$ and complexity $C$ (the set of features). To perform the classification, data is randomly partitioned into $n$ equally sized samples that preserve the total fraction of occurrences in each category. One of the samples is used for validating the algorithm, and the remaining $n-1$ samples are used for training the algorithm. The accuracy (that is, the fraction of correct identifications) obtained from the training set is the training score, and the one obtained from the validation set is the cross-validation score. This process is repeated $n$ times, producing an ensemble from which the average values of the scores and their confidence intervals are calculated. This approach is known as the $n$-fold cross-validation strategy~\cite{hastie2013elements}.

We estimate the training and the cross-validation scores for each machine learning algorithm as a function of their main parameters (the validation curves). This is common practice for estimating the best trade-off between bias and variance errors. Bias errors occur when the learning methods are not properly taking into account all the relevant information about the explanatory variables that describe the data (underfitting). Variance errors, on the other hand, usually happen when the complexity of the learning model is too high, that is, high enough even for modeling the noise in the training set (overfitting). SI Appendix, Fig.~S10 shows the validation curves for the four learning methods that we use in our study. The parameters that we have studied are the number of neighbors in the case of the $k$-nearest neighbors algorithm, the number and the maximum depth of trees in the case of the random forest method, the parameter associated with the width of the radial basis function kernel and the penalty parameter for the support vector machine classification, and the so-called $L2$ penalty for the neural network model. We use these results as a guide for applying a more exhaustive grid-search algorithm~\cite{scikit-learn}, from which the best tuning parameters of each learning method are obtained. These values are reported in SI Appendix, Fig.~S10 and used for obtaining the results shown in~\autoref{figure4}C.

In addition to the validation curves, we have also estimated the learning curves, that is, the dependence of the training and the cross-validation scores on the size of the training set. This practice is also common when dealing with statistical learning algorithms, since very small training sets are usually not enough for fitting the model, while adding unnecessary data may introduce noise to the model. The results presented in SI Appendix, Fig.~S10 show that the cross-validation score increases with the training size for all algorithms. However, this growth is practically not significant if the training set exceeds 50\% of the data.

}

\showmatmethods{} 

\acknow{This research was supported by CNPq, CAPES (Grants Nos. 440650/2014-3 and 303642/2014-9), and by the Slovenian Research Agency (Grants Nos. J1-7009 and P5-0027).}

\showacknow{} 


\bibliography{complexity_paintings}
\end{document}